# Building Structures Atom-by-Atom via Electron Beam Manipulation

Notice to the editor (not to be published): This manuscript has been authored by UT-Battelle, LLC, under Contract No. DE-AC05-00OR22725 with the U.S. Department of Energy. The United States Government retains and the publisher, by accepting the article for publication, acknowledges that the United States Government retains a non-exclusive, paid-up, irrevocable, world-wide license to publish or reproduce the published form of this manuscript, or allow others to do so, for United States Government purposes. The Department of Energy will provide public access to these results of federally sponsored research in accordance with the DOE Public Access Plan (http://energy.gov/downloads/doe-public-access-plan).

# Building Structures Atom-by-Atom via Electron Beam Manipulation

*Ondrej Dyck[*,1,2], Songkil Kim[3], Elisa Jimenez-Izal[4,5], Anastassia N. Alexandrova[5,6], Sergei V. Kalinin[1,2], Stephen Jesse[1,2]*

[1]The Center for Nanophase Materials Sciences, Oak Ridge National Laboratory, Oak Ridge, Tennessee, 37831, USA

[2]The Institute for Functional Imaging of Materials, Oak Ridge National Laboratory, Oak Ridge, Tennessee, 37831, USA

[3] School of Mechanical Engineering, Pusan National University, Busan 46241, South Korea

[4] Department of Chemistry and Biochemistry, University of California, Los Angeles, Los Angeles, California, 90095-1569.

[5] Kimika Fakultatea, Euskal Herriko Unibertsitatea (UPV/EHU), and Donostia International Physics Center (DIPC), P. K. 1072, 20080 Donostia, Euskadi, Spain

[6] California NanoSystems Institute, 570 Westwood Plaza, Building 114, Los Angeles, CA 90095.

ABSTRACT

Building materials from the atom up is the pinnacle of materials fabrication. To date the only platform that offers single atom manipulation is scanning tunneling microscopy. Here we demonstrate controlled manipulation and assembly of few atom structures by bringing together single atoms using a scanning transmission electron microscope. An atomically focused electron beam is used to introduce Si substitutional defects and defect clusters in graphene with spatial control of a few nanometers and enable controlled motion of Si atoms. The Si substitutional defects are then further manipulated to form dimers, trimers and more complex structures. The dynamics of a beam induced atomic scale chemical process is captured in a time-series of images at atomic resolution. These studies suggest that control of the e-beam induced local processes offers the next step toward atom-by-atom nanofabrication, providing an enabling tool for the study of atomic scale chemistry in 2D materials and fabrication of predefined structures and defects with atomic specificity.

INTRODUCTION

Fabrication of matter atom by atom remains a long-standing dream and ultimate goal of nanotechnology, following the famous challenge by Feynman 58 years ago.[1] For 30 years, the atom by atom fabrication remained the province of visionary thinking and science fiction, inspiring but seemingly unachievable given the then available fabrication tools. The situation changed drastically upon the introduction of scanning tunneling microscopy (STM) by Binnig and Rohrer[2-4] and subsequently STM-based atomic fabrication by Don Eigler.[5-8] This advancement immediately riveted the attention of both the scientific and general community worldwide, launching the era of nanotechnology. It also started the developments of technologies based on combined STM and surface science methods, ultimately leading to the development of single-atom qubit devices.[9-11]

Despite the remarkable progress in STM based atomic fabrication, the fabrication process remains slow and requires complex surface science approaches to establish and maintain atomically clean surfaces. This process further requires a complex technological cycle to integrate single-atom devices with the classical semiconductor technologies. Correspondingly, the development process remained slow and required large capital investments to even begin the development. Thus, alternative methods for atom-by-atom fabrication are of interest.

In recent years, the ever-growing body of work in high resolution scanning transmission electron microscopy has illustrated the potential of the electron beam to induce local atomic-scale changes in materials microstructure that can be immediately visualized.[12-19] Almost immediately, it was proposed that the e-beam can be used for fabrication of atomic-scale structures.[20-28]

Following these predictions, Susi et al[18,29] demonstrated controllable e-beam-induced movement of a single Si a short distance through the graphene lattice as well as clarifying the mechanisms of Si motion in graphene and calculating the energy required for various processes involved in atomic motion and the creation of point defects. This built upon previous, related work,[13,19,30,31] and was expanded upon by additional studies by Robertson et al[32-36] and others[17,37-39] which served to clarify various atomic scale processes and transformations in graphene. Dyck et al[40] has further shown an approach to incorporate single Si dopants into the graphene lattice at preselected locations, localizing single dopant atoms at specific lattice sites. Based on these studies, we aim to further develop atomic scale control and, critically, demonstrate atom-by-atom assembly by electron beam manipulation. Here we demonstrate the transition from single dopant control towards the creation of a Si dimer, trimer, and tetramer from Si substitutional point defects.

MATERIALS AND METHODS

Graphene was grown on a Cu foil via atmosphere pressure chemical vapor deposition (APCVD).[41] The Cu foil was spin-coated with poly(methyl-methacrylate) (PMMA) to form a mechanical stabilizing layer, after which the foil was dissolved away in an ammonium persulfate-deionized (DI) water solution (0.05 g/ml). The graphene/PMMA film was transferred first to hydrogen chloride (HCl) diluted in DI water and then to a DI water bath for the removal of ammonium persulfate residue, followed by a final rinse in a fresh DI water bath before being transferred to a TEM grid. The graphene/PMMA film was then scooped from the bath with a TEM grid and heated on a hot plate at 150 °C for ~20 min to make better adhesion to the grid substrate. Then, most of the PMMA was dissolved in an acetone bath, followed by an isopropyl alcohol rinse. Finally, the grid was annealed in an Ar-$O_2$ (450 sccm/45 sccm) environment at 500 ºC for 1.5 hours to mitigate hydrocarbon deposition in the microscope.[42,43]

Imaging of the sample was performed in a Nion UltraSTEM U100 at an accelerating voltage of 100 kV and 60 kV, as indicated in the text. At 100 kV the beam current was 30-40 pA and at 60 kV the beam current was 60-70 pA. The medium angle annular dark field (MAADF) detector was used for imaging at 100 kV to enhance image contrast and the high angle annular dark field (HAADF) detector was used for imaging at 60 kV in order to preserve Z-contrast.[44]

COMPUTATIONAL METHODS

Geometry optimization of defected graphene was carried out within density functional theory (DFT), with the Perdew–Burke–Ernzerhof (PBE)[45] exchange-correlation functional and the projected augmented wave (PAW) method,[46,47] as implemented in the Vienna *ab-initio* simulation package (VASP).[48] We used a plane-wave kinetic energy cutoff of 520 eV and $\Gamma$ centered 4x8x1 k point grid. The convergence criterion was set to $10^{-5}$ ($10^{-6}$) eV for geometry (electronic) and 0.02 eV/Å for forces. The DFT-D3 with Becke-Jonson damping scheme[49,50] was used to account for the dispersion interactions. We first optimized the lattice of perfect graphene, resulting in a lattice constant of 2.468 Å, and a C-C bond length of 1.425 Å. Defected graphene was then modeled using an 8x4 unit cell, containing a total of 64 C atoms. For each Si doped structure modeled in the present work, not only planar geometries were considered, but Si atoms were placed below and above the plane in all the possible configurations. A more accurate $\Gamma$ centered 12x24x1 k point grid was used for the projected density of states (PDOS).

INTRODUCING ATOMIC SCALE DEFECTS

The first step of e-beam fabrication of atomic structures in graphene is the controllable introduction of substitutional defects within the pristine graphene lattice. Previously, we demonstrated how single atoms or small clusters of atoms may be introduced into a graphene lattice.[40] This approach has the advantage of highly precise positioning of the defect in the lattice. However, it is fairly slow, taking a couple of minutes per defect, and risks damaging the graphene lattice with the 100 keV beam to the point that it is unable to heal. Thus, it becomes somewhat tedious to control beam position, blanking, scanning, and imaging in such a way as to produce single defects with the care required to protect the rest of the graphene from beam damage.

In an attempt to develop a more user-friendly way of introducing point defects into the lattice we developed two alternative techniques which sacrifice the precise positioning of the previous technique but gain in ease of execution. Figure 1 summarizes these two techniques. (a)-(d) show the first technique. In (a), a region (circled) is chosen near the source material. The e-beam is then drawn via operator control from the source material onto the pristine graphene lattice a few times (on the order of 1-3 seconds of total exposure time to the 100 keV electron beam), (b). This causes a repeated agitation and sputtering of the source material followed by the introduction of vacancies in the graphene lattice. This procedure decreases the likelihood of creating a large hole in the graphene that will not heal, because the beam is not left stationary. In this approach, the location of the defects created are spread over 1-2 nm in the locations over which the beam was moved. While this method appears to work fairly consistently and easily, there remains the danger of accidentally introducing large holes that will not heal. The main reason is that while the beam is under manual control in this way, imaging is not possible so real-time monitoring of the state of the sample is precluded and large holes may still be formed accidentally.

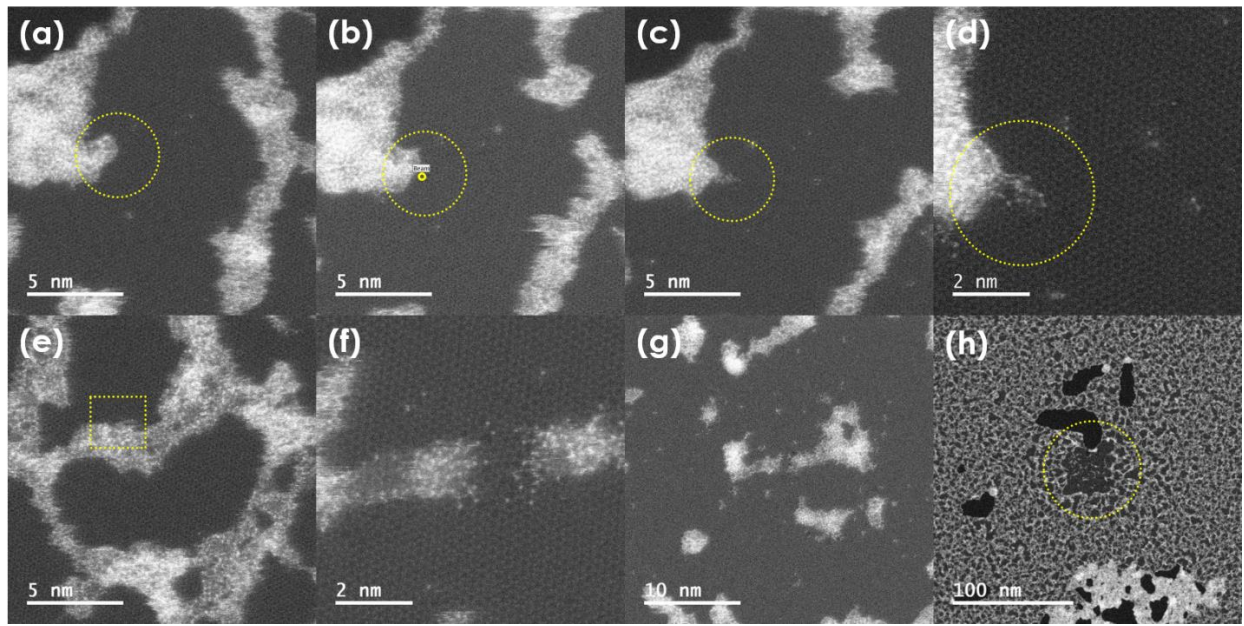

**Figure 1**. Introduction of Si substitutional defects and defect clusters into a graphene lattice (a)-(d). An area close to the source material is chosen, (a). The stationary beam is moved "by hand" over areas where defects are to be created, (b). The small circle indicates the beam location, which was moved back and forth from the source material to the graphene. As atoms from the source material were introduce into the lattice we begin to see a bright area where the beam was moved, (c). A magnified view is shown in (d) where the graphene lattice and defect clusters are easily observed. (e)-(h) show a similar but distinct way of introducing many substitutional defects from the source material. A typical area of the source material is shown in (e). A sub-scan area is selected, as indicated by the box, and manually moved along the edges of the source material so that the graphene lattice and the source material is irradiated. This simultaneously sputters away the source material and creates vacancies in the graphene lattice which have a high likelihood of becoming passivated by a sputtered Si atom. A typical state after a few seconds of scanning, (f), where many defects have been formed in the graphene lattice and much of the source material is gone. (g) shows the same area as (f) but at a lower magnification. (h) shows a much lower magnification of the irradiated area, circled, and the surrounding web of source material.

Figure 1 (e)-(h) illustrates an alternative technique for the introduction of substitutional defects within a defined area. With this method, an overview image is acquired, (e), and a sub-scan location is selected over both the pristine graphene and source material, illustrated with the box. The sub-scan location may be moved around via operator control while the beam is scanning within the subscan region. The image produced from the sub-scan area can then be used to monitor the state of the sample in real time. This procedure also both sputters the source material and introduces vacancies in the graphene lattice. The results of this procedure are shown at various magnifications in (f)-(h). In (f) individual substitutional Si atoms have been introduced into the graphene lattice. While it is difficult to measure exact exposure times to a specific area due to the dynamic motion of the subscan box, exposures on the order of 10-30 s are typical. A lower magnification image is show in (g) where many defects and small clusters of defects can be seen stuck in the lattice. Finally, (h) is further demagnified to contrast the final state of the region of interest with the surrounding area. Much of the source material has been sputtered away revealing tens of nanometers of mostly pristine graphene harboring the introduced defects.

EXTENDED MOVEMENT OF SI DOPANT ATOMS

Once the substitutional Si defects have been introduced, we can begin to explore the capabilities of a STEM for atomic level control. Movement of single Si atoms through a graphene lattice has been explored previously.[18,29,39] A critical enabling aspect of atomic scale manipulation in STEM will be to establish sample treatment procedures that enable extended movement of the introduced dopant atoms. To this end, here we show an example of directionally controlled atomic motion over a path length of 4.5 nm achieved on a sample using the treatment procedure described. The accelerating voltage used here was 60 kV with a convergence angle of 30 mrad and beam current of ~ 60 pA. **Figure 2** shows a summary of the directed motion. The images were acquired with the minimum dose necessary to distinguish the lattice. They were artificially colored with the fire look up table in ImageJ and blurred with a Gaussian to aid in visibility. In a similar experiment we moved a Si atom in a circle to prevent it from moving out of the field of view and were able to achieve a total path distance of 6.5 nm. Videos of both are provided in the supplementary information. Thus, we conclude that the sample treatment procedure has allowed imaging and manipulation over extended distances without unwanted contamination or chemically reactive elements altering the defects.

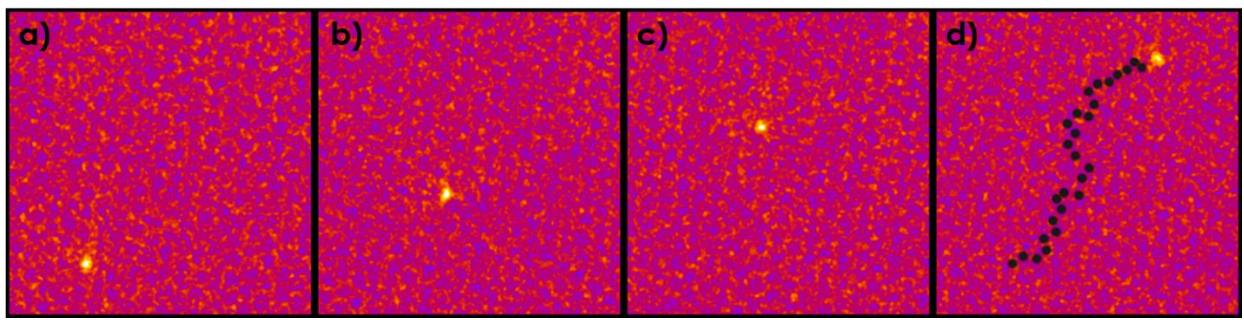

**Figure 2.** Illustration of beam-induced atomic motion over a path length of 4.5 nm. The focused electron beam was positioned on top of adjacent C atoms to encourage the bright Si dopant to exchange places with the irradiated neighbor. a) shows the initial configuration and b)-d) show subsequent images of the progress. d) indicates the recorded lattice positions of the Si atom through time (black dots). A video of each acquired frame is presented in the supplementary information. The field of view is 4 nm.

FORMATION OF SILICON DIMERS

The ability to move single dopant atoms suggests that the construction of structures atom-by-atom is achievable via *in situ* STEM and, here, we explore a few simple structures fabricated in this way. The results show in **Figure 1** were acquired using a 100 keV electron beam to assist in the production of defects in the graphene lattice. For the rest of the experimental results, the accelerating voltage was lowered to 60 kV to prevent continued damage and allow closer inspection of the formed defects.

**Figure 3** (a) shows a HAADF image of two Si substitutional point defects introduced into the lattice within close proximity of one another. Both defects exhibit three-fold coordination with the carbon lattice. A small sub-scan area was used to direct the electron beam onto carbon atoms adjacent to the Si atoms in the desired direction of motion. A short video of the directed motion of a Si atom performed in this manner may be found in the supplemental materials. In this example, we attempted to move the upper left Si atom. The circle marks the starting location in (b)-(g) and the dotted lines record the atom's movements through time. In (d), the process of acquiring the image resulted in the unintentional movement of the lower left Si atom. In (f) the sub-scan area was large enough to cause both atoms to move to the positions shown in (g). The inset in (g) shows an atomic model of the configuration at this stage. The two Si atoms are sitting opposite each other in the hexagonal ring. An attempt was made to pull them closer together by scanning over the two carbon atoms at the top of the ring, but instead, a Si atom was ejected from the lattice and the material restructured into a single Si occupying two lattice sites in the four-fold configuration, (h). Given that the majority of the beam fluence was focused onto the adjacent carbon atoms, it is likely that one of the Si atoms transitioned to one of these sites first and then, being under the beam, was ejected from the lattice. The inset in (h) shows an atomic model of the final configuration.

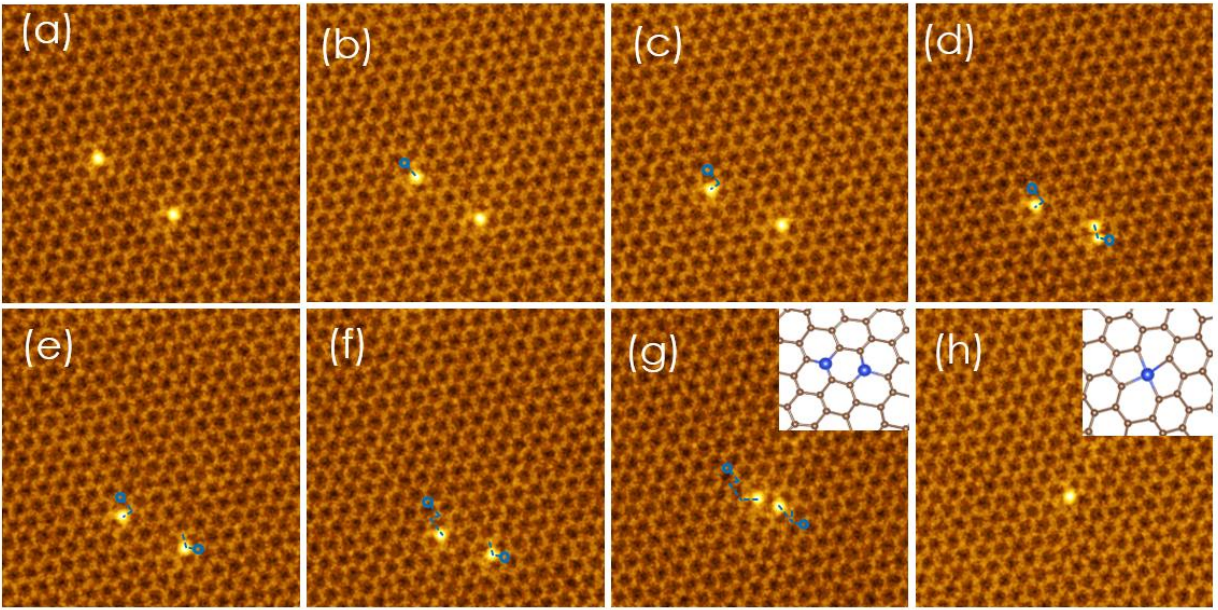

**Figure 3.** Creation of a Si dimer within a graphene lattice via e-beam manipulation. (a) shows the starting configuration with two 3-fold coordinated Si atoms sitting in substitutional lattice sites. In (b)-(g) the motion of the two Si atoms are tracked as they are moved. The original positions are indicated by the small circles and the dotted line records the atom positions through time. Between (g) and (h) a Si atom was ejected from the lattice (unintentionally) and the remaining Si atom became 4-fold coordinated to occupy two lattice sites, accounting for the loss of an atom from the lattice.

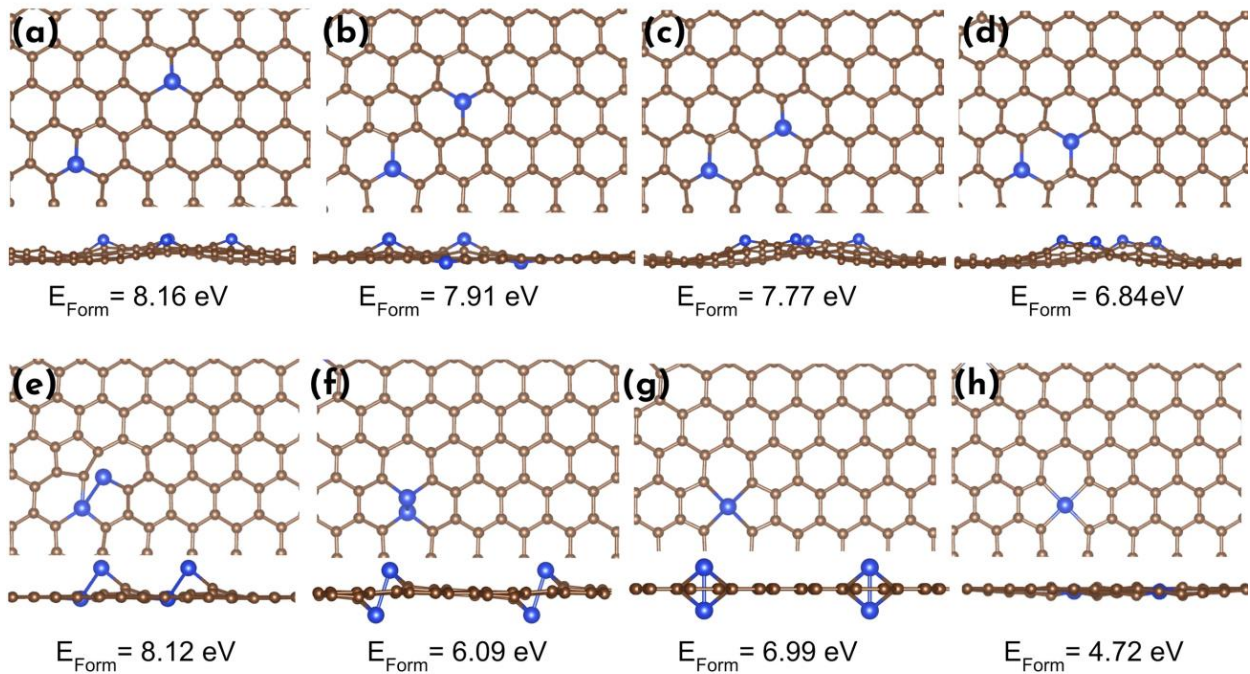

**Figure 4** Final structures optimized with DFT. Si atoms are approaching each other from (a) to (g). For each structure a top-down view and lateral view are shown, along with the formation energy, in eV. Carbon atoms are depicted in brown and Si atoms in blue.

In order to understand more clearly the formation and stability of the structures formed in **Figure 3**, we performed DFT calculations of similar structures (shown in **Figure 4**), exploring the relaxed configurations and energies of two Si substitutional defects as they approach each other in the graphene lattice. We modeled eight configurations, starting from a Si-Si distance of 6.56 Å (**Figure 4** (a)), and decreasing the distance between them by one lattice site for each new configuration, (b)-(d). In (e)-(g) we modelled three stable configurations where the Si atoms are in closer proximity to each other than the structure shown in (d). These structures were not observed in the experiments performed here but, nevertheless, represent possible stable structures that might be observed (see for example Susi et al[18] supplementary information). The final structure observed experimentally is shown in (h), with one Si atom in a 4-fold coordination. The optimized monolayers evidence a clear preference for corrugated structures, to the detriment of completely planar ones. This can be easily understood considering that silicon is bigger than carbon. Indeed, we find that in general, the closer the Si atoms are, the less stable becomes the planar configuration. For (f) and (g) no planar variations are stable. The formation energies of the defects were calculated as $E_{Form} = E_{def} + n\mu(C) - m\mu(Si) - E_{perfect}$, where $E_{def}$ and $E_{perfect}$ are the energy of the defective and perfect graphene respectively, $\mu(C)$ is the chemical potential of C, estimated as the total energy per atom in graphene, $\mu(Si)$ is the chemical potential of Si, calculated from the total energy per atom of Si in the bulk Si,[51] and *n* and *m* are the total number of C atoms removed and number of Si atoms added to the defected structure. The formation energies reveal the preference for Si atoms to be close to each other, in agreement with the results of Susi et al.[18] presented in the supplemental information. The formation energy decreases when going from (a) to (d), i. e., when the two Si atoms approach each other. $E_{form}$, however, goes up in (e) and reaches its minimum value for (h), with just one silicon atom 4-fold coordinated in the lattice. This correlates well with the experiment, since structures (e)-(g) were not observed. The relative energy for structures (d) and (h) as calculated in reference 18, is $E_{rel} = E^h_{form} - E^d_{form} = 2.12$ eV, which means that the structure with just one 4-fold coordinated Si atom (h) is 2.12 eV more stable than the structure with two Si atoms close to each other (d). It is worth commenting on the possibility of structures (f)-(h) being that observed in **Figure 3** (h). It is possible that the Si atoms in structure (f) would be too close together to resolve in the experiment and may appear to be just a single atom, but in this configuration, we would see an elongated bright spot in the experimental image, which does not appear to be the case. Conversely, if the configuration in (g) was captured, where the Si atoms stack on top of the other, the Si atoms would become brighter than a single Si atom. Again, this is not what we see in the experimental images. Thus, we conclude that we are observing the configuration shown in (h). Considering that we are manipulating the sample with a 60 keV electron beam, it is possible that one of the two Si atoms gained enough energy to escape from the lattice. Furthermore, due to the great affinity of silicon with oxygen, it is also possible that foreign adatoms may have chemically facilitated this process. We additionally calculated the formation energy of a similar structure to (h), but with a second Si placed at ~12 Å from the lattice, representing e-beam induced ejection from the lattice into the vacuum. This energy is 10.58 eV and can be viewed as an upper limit required for this transition given that it is likely that the ejected atom remains a loosely bound adatom which quickly diffused away from the imaged area[52] or was involved in an additional chemical reaction with, for example, oxygen or hydrogen.

Given that Si atoms in the graphene lattice have two frequently observed coordinations, we further assembled two four-fold coordinated Si substitutional defects into a dimer with a different final structure. **Figure 5** shows a process where two four-fold coordinated Si atoms were brought together via electron beam manipulation by performing a sub-scan over the carbon atoms between the two Si atoms, (a)-(c). We remind the reader that the four-fold coordinated Si atoms replace two C atoms and create two adjacent pentagonal rings. As they are brought together in (c), a pentagonal ring from each of the Si defects merge to form a 5-8-5 structure.

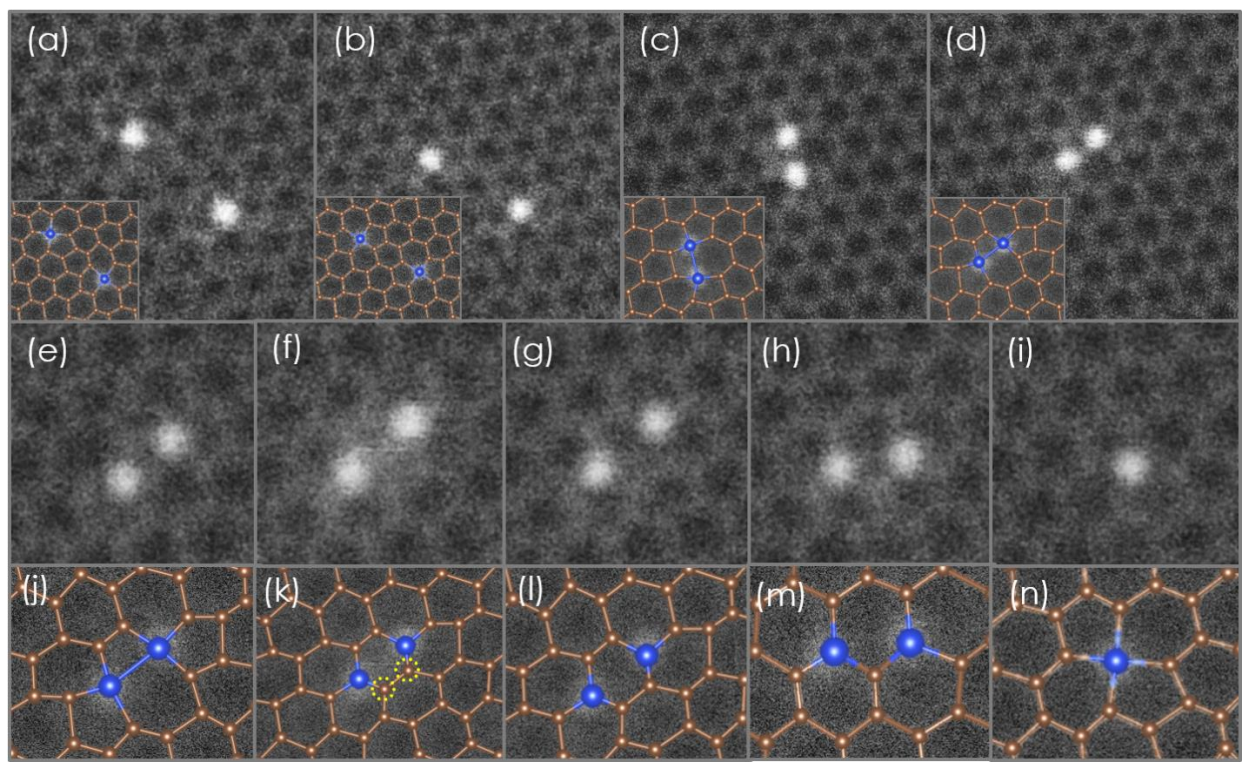

**Figure 5** Creating an alternately structured Si dimer and observing an atomic scale chemical reaction. (a)-(c) show the result of pulling together two four-fold coordinated Si atoms. (d)-(i) show the evolution of the created dimer under continued e-beam irradiation. (j)-(n) show atomic models overlaid on the images from (e)-(i). The atomic models are based on the images only and do not represent theoretical modelling. (e)-(i) are frames taken from a ten-frame video which may be found in the supporting information. During the acquisition of the video two additional carbon atoms are introduced into the structure, circled in (k). This converts the Si atoms to three-fold coordination. The right-hand Si atom moves down in (h) and (m) and, finally, the left-hand Si atom is ejected from the lattice in (i) and (n), resulting in a single four-fold coordinated Si. From (c) to (d), the electron beam irradiation induced rotation of the defect. This is similar to the observations of Yang et al.[37]

The images shown in (e)-(i) are frames from a ten-frame video (included in the supplemental information) with (j)-(n) showing the corresponding atomic models. The starting configuration, (e), is the same as that shown in (d). In (f) there occurred a slight blur around the Si atoms as two C atoms flew into the illuminated region and restored the lattice to hexagonal symmetry. The two new atoms are indicated by dotted circles in the atomic model, (k). In (g), the new C atoms stopped moving, the blur disappeared, and the lattice was converted to the same structure as that shown in **Figure 3** (g), namely two three-fold coordinated Si atoms sitting opposite each other in a hexagonal ring. In (h) and (m), the right-hand Si atom moved down one lattice site and in (i) and (n) the left-hand Si atom was ejected from the lattice. The ejection of the one Si atom requires that the remaining Si atom assume a four-fold coordination to occupy two lattice sites.

FORMATION OF SI TRIMER AND TETRAMER

As a final example of e-beam assembly of nanostructures embedded in graphene we illustrate formation of a Si trimer and tetramer from a dimer of the same configuration as that formed in **Figure 5** (a)-(c). The dimer shown in **Figure 6** (a) was introduced into the graphene lattice via the process described at the beginning, whereby we created a host of defects in the graphene lattice. This is the same defect structure observed upon assembling two four-fold coordinated Si substitutional defects. In **Figure 6** (b) a bond rotation was induced by the scanning electron probe involving the carbon atoms adjacent (below) the Si dimer, arrowed in the figure. This rotation likely occurs on the order of ~100 fs (see Yang et. al.[37] supplemental information and Susi et. al.[18] supplemental video). This is 11 orders of magnitude faster than our pixel dwell time and therefore not imageable in transition. Because the imaging process is extended through time, what we are observing is the juxtaposition of the initial configuration in the top half of the defect, before the rotation, and the final configuration on the bottom half, after the rotation. In (c) the bond rotation has completed. In (d) we observe a mobilized Si adatom has attached itself to the defect. We address how this may be accomplished later. This adatom was momentarily knocked away by the beam as shown in (e). Upon reattaching to the implanted defect, we observe a beam-induced exchange of the two carbon atoms involved in the rotation with the Si adatom, the final configuration of which is shown in (f). Yang et. al.[37] also observed this bond rotation and acquisition of a third Si atom and provide a more in-depth discussion, which we will not repeat here. We merely repeat that this trimer can be rotated easily with the electron beam positioned on top of a carbon atom adjacent to the trimer, as shown in (g). To attach a fourth Si atom to the structure, (h), the beam was scanned over a large area (~50-100 nm) to agitate and mobilize surrounding source material and adatoms. Within a few tens of seconds, the fourth Si atom was added to the structure. We suggest this method will also reproducibly attach a third Si atom to the dimer, as observed in d. This new structure could also be rotated, shown in (i), though not as freely. Positioning the beam on the central Si atom lead to the ejection of this atom from the structure and a return to the trimer configuration, shown in (j). All configurations shown in (g)-(j) could be repeatably brought about through controlled e-beam exposure. It is less clear, however, weather the bond rotation and subsequent capture and incorporation of a Si adatom is the most favorable formation pathway for the trimer structure (we have also observed trimers and tetramers formed during the dopant insertion stage). Nevertheless, given that Yang et al[37] observed the bond rotation to be reversible, and that the capture of the third Si atom can be intentionally directed, as evidenced by the capture of the fourth Si atom, we conclude that each of these structures can be formed through careful e-beam exposure.

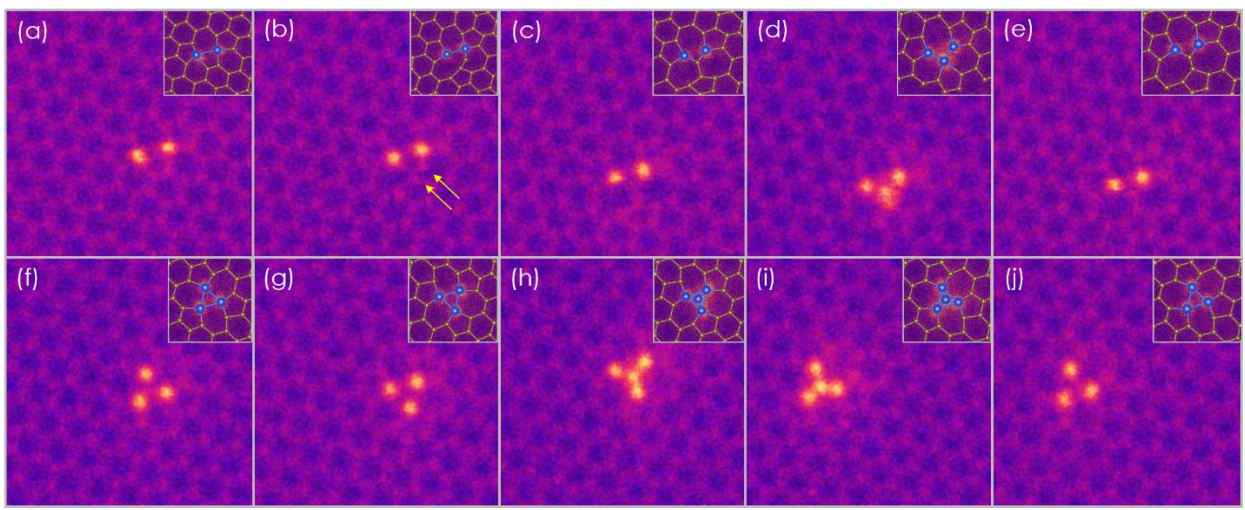

**Figure 6** Evolution of a Si dimer under the influence of the 60 keV beam. Insets show the suggested atomic models based on the images (not simulation). (a) shows the initial configuration. In (b) we observe a bond rotation beginning to occur next to the Si dimer (arrowed). (c) shows the configuration after the bond rotation has occurred. In (d) we observe a Si adatom temporarily attaching to the defect. The adatom was knocked away and we return to the configuration shown in e) which appears identical to that in (c). In (f) the adatom is recaptured and incorporated into the lattice. Once in this configuration, all subsequent configurations, (g)-(j), could be repeatably produced through electron beam manipulation. Images were artificially colored using the fire look up table in ImageJ.

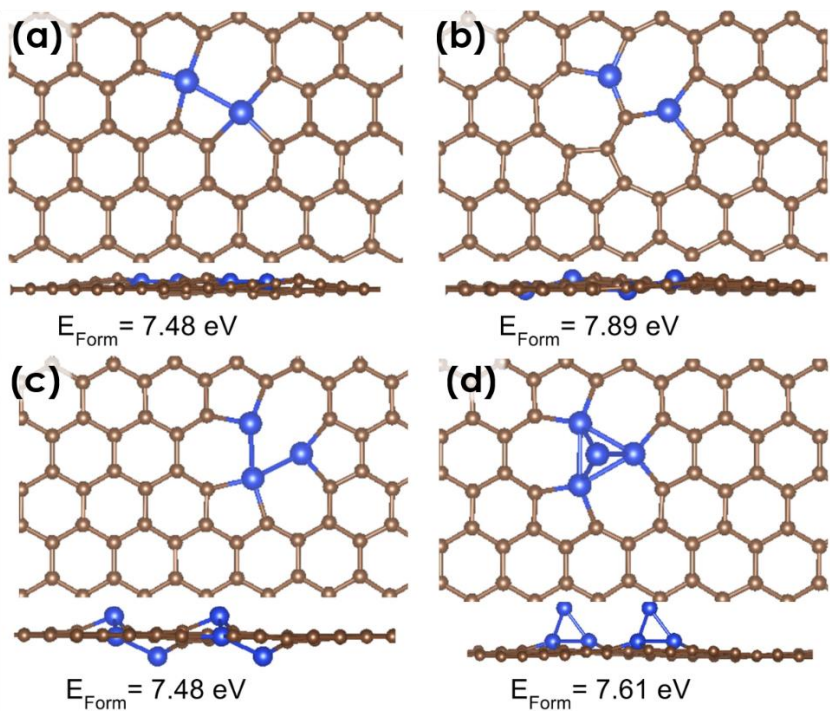

**Figure 7** Calculated most stable structures observed experimentally in **Figure 6Figure 6**. Carbon atoms are depicted in brown and Si atoms in blue.

Using DFT we modeled various Si doped graphene monolayers found in **Figure 6**. The optimized structures are shown in **Figure 7**, along with the calculated formation energies. Note that from **Figure 7** (b) to (c) the two C atoms involved in the bond rotation were lost. For the Si trimer, **Figure 7** (c), we found that one Si sits in-plane, a second one below and a third above the plane, in agreement with the findings of Yang et al..[37] In the tetramer, **Figure 7** (d), the central Si is above the rest of the three Si atoms at a 2.38 Å distance. This is the only stable geometry we found for the tetramer observed experimentally. Interestingly, the energy required to create the Si trimer is the same as for the Si dimer in (a) and it is similar to the energy required to create the tetramer. Finally, the partial density of states (PDOS) plots built for a variety of Si doped graphene monolayers (see supplementary information) show that the $p_z$ orbitals are closer to the valence band edge.

CONCLUSION

We have shown how Si substitutional defects and Si defect clusters may be introduced into a graphene lattice via in situ STEM techniques. A prerequisite for this process is that the samples remain free of e-beam induced hydrocarbon deposition. This is achieved through an $ArO_2$ annealing process which we borrowed from Garcia et al[43] and investigated previously.[42] We demonstrate that directing a 100 keV focused electron beam across the source material and the graphene lattice can reliably generate multiple substitutional Si defects. Subsequent manipulation of the introduced defects may be accomplished by decreasing the microscope accelerating voltage to 60 kV and using controlled scan areas or direct stationary beam irradiation to induce movement. We showed controllable e-beam induced formation of two Si dimers from two three-fold coordinated Si substitutional atoms and two four-fold coordinated Si substitutional atoms. We also documented a formation pathway for Si trimers and tetramers, reversible conversion of a Si trimer to tetramer, and the controllable rotation of both structures. DFT modelling was performed to more clearly understand the structures formed, the energies required to form them, and explore other possible stable structures. These represent the first steps toward general atomic scale manufacturing.

Finally, we also captured an image time-series of an atomic scale chemical reaction occurring with atomic resolution. As technologies such as detector efficiency, compressed sensing, and artificial intelligence improve, such observations will become more commonplace and will provide a wealth of information to enhance our understanding of atomic scale physics and our mastery of materials. In the words of Feynman "I am not afraid to consider the final question as to whether, ultimately---in the great future---we can arrange the atoms the way we want; the very atoms, all the way down! What would happen if we could arrange the atoms one by one the way we want them."[1] This great future is unfolding before us.

ACKNOWLEDGMENT

We would like to thank Dr. Ivan Vlassiouk for provision of the graphene samples and Dr. Francois Amet for assistance with the argon-oxygen cleaning procedure.

Research was performed at the Center for Nanophase Materials Sciences, which is a US Department of Energy Office of Science User facility. Experimental work was supported by the Laboratory Directed Research and Development Program of Oak Ridge National Laboratory, managed by UT-Battelle, LLC for the U.S. Department of Energy (O.D., S.K., S.V.K., S.J.).

SUPPORTING INFORMATION

Additional information regarding the DFT simulations (file type, PDF)

Video clip of moving a Si atom (file type, AVI)

Video clip of moving a Si atom over several nanometers (file type, AVI)

Video clip of moving a Si atom in circles to prevent it from leaving the field of view (file type, AVI)

Video clip of an atomic scale chemical reaction (file type, AVI)

| Structure | ΔE (eV) |
|---|---|
| **1** | 1.62 |
| **2** | 1.81 |
| **3** | 1.77 |
| **4** | 2.37 |
| **5** | 1.31 |
| **6** | - |
| **7** | - |

**Table 1** Energy difference between the structures shown in **Figure 3** and the corresponding planar structures, where all the atoms lie on the same plane, calculated as $\Delta E = E_{buck} - E_{planar}$, where $E_{buck}$ is the energy of the buckled structure and the $E_{planar}$ the energy of the planar structure. Energies in eV. For the configurations **6** and **7** planar structures are unstable.

| Structure | ΔE (eV) |
|---|---|
| **A** | 0.76 |
| **C** | 0.02 |
| **D** | 1.23 |
| **D'** | 1.23 |
| **H** | - |

**Table 2** Energy difference between the structures shown in **Figure 6** and the corresponding planar structures, where all the atoms lie on the same plane, calculated as $\Delta E = E_{buck} - E_{planar}$, where $E_{buck}$ is the energy of the buckled structure and the $E_{planar}$ the energy of the planar structure. Energies in eV. The tetramers, **H**, do not form planar structures.

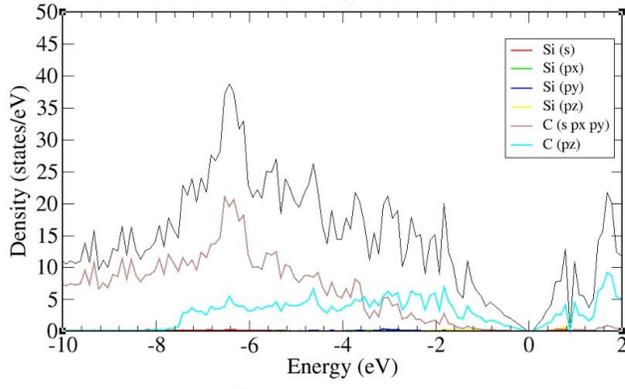
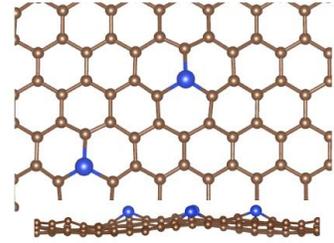
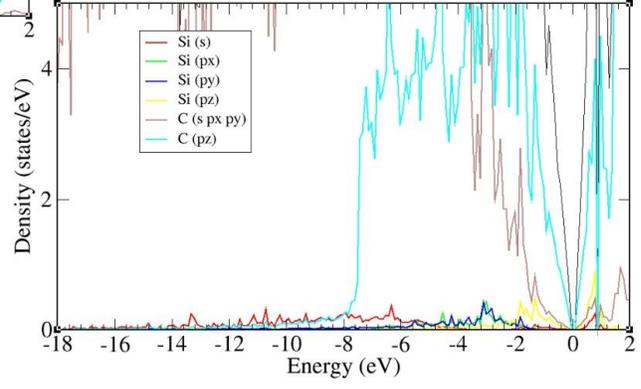
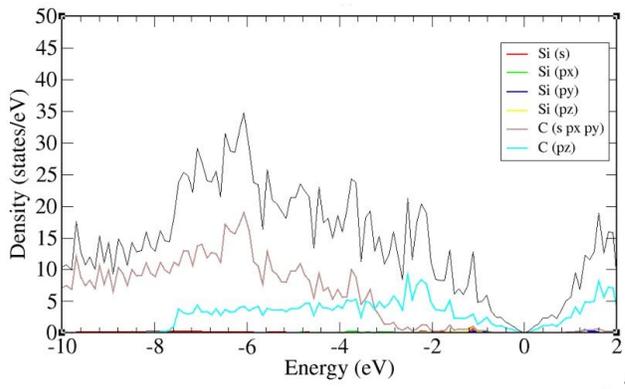
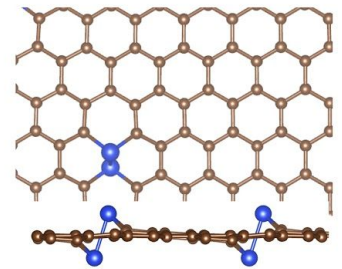
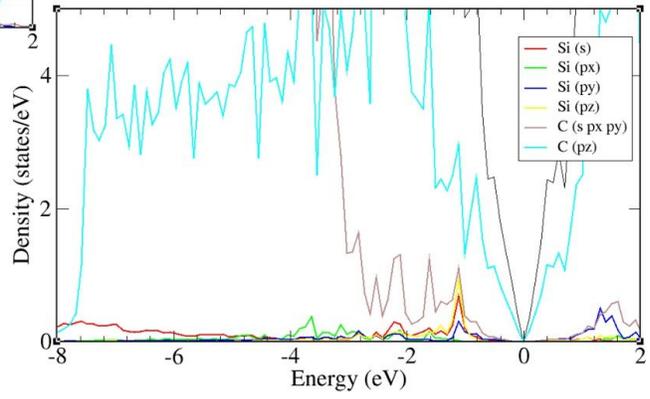

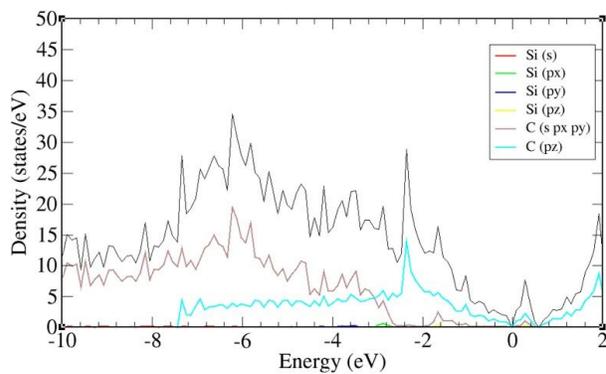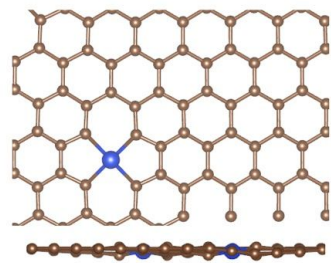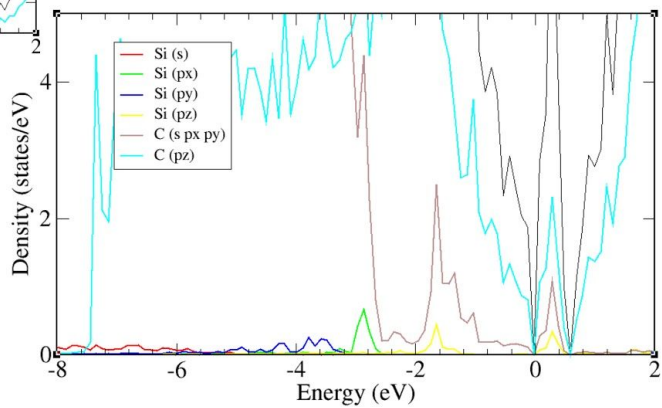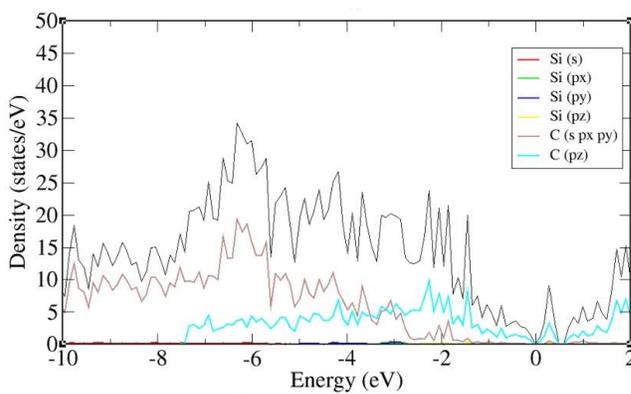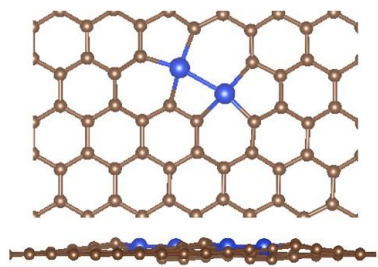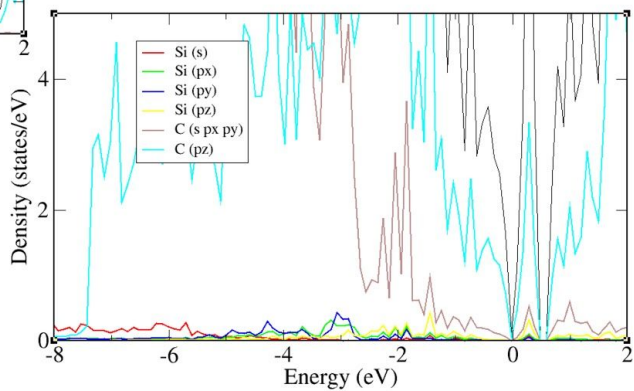

ZOOM

ZOOM

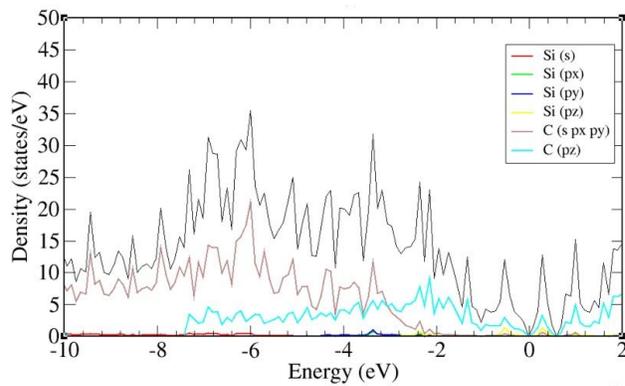
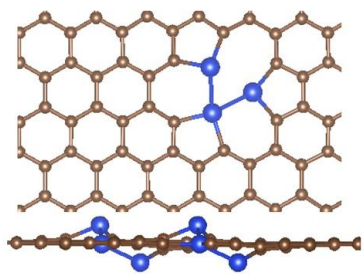

D

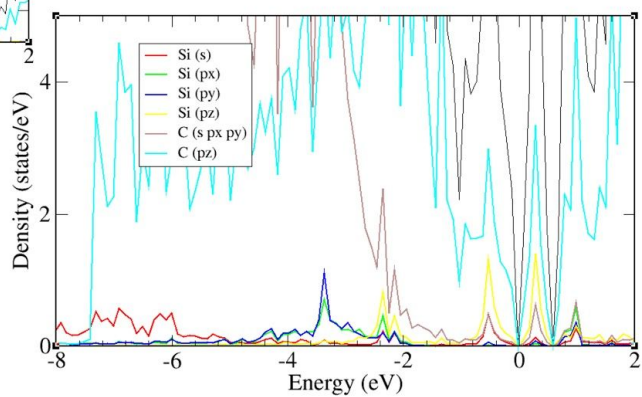

ZOOM

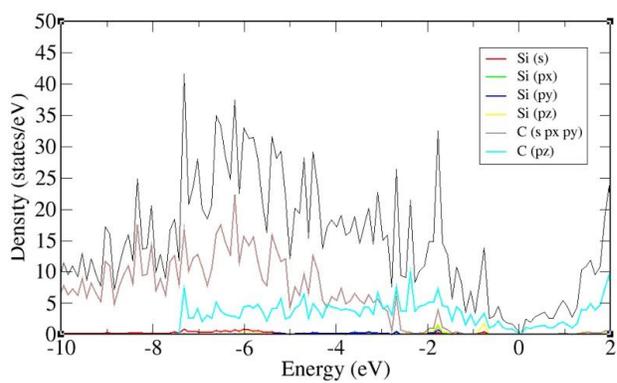
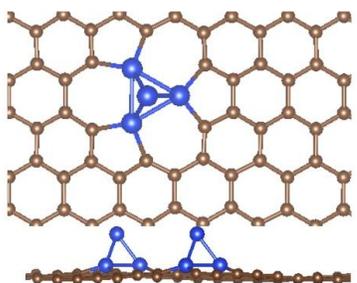

H

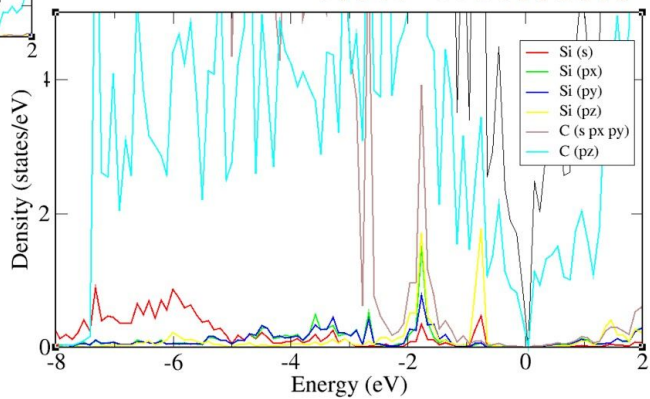

ZOOM

**Figure S1** Projected density of states of various Si doped graphene structures. Γ-centered 12x24x1 k point grid was used for these images.